\documentclass[aps,twocolumn,showpacs,amsmath,amssymb,floatfix]{revtex4-1}

\usepackage{graphicx}
\usepackage{amsmath,bm,amssymb,amsfonts}

\begin{document}

\title{Spin Hall effect in Rashba-Dresselhaus planar strips in the
presence of electron correlations}

\author{Jos\'e A. Riera}
\affiliation{Instituto de F\'{\i}sica Rosario, Consejo Nacional de
Investigaciones Cient\'ificas y T\'ecnicas, and
Universidad Nacional de Rosario, Rosario, Argentina
}

\date{\today}

\begin{abstract}
A model with both Rashba and Dresselhaus spin orbit (SO) couplings
and Hubbard electron-electron interaction is studied on planar strips
at quarter filling at zero temperature in the clean limit. In the
absence of Hubbard repulsion and at equilibrium, within linear
response theory, a nonmonotonic behavior of the spin Hall conductivity
as a function of the ratio of the Rashba ($\alpha$) and Dresselhaus
($\beta$) strengths was found for large enough SO strengths.
This behavior is signalled by a peak or a cusp, depending on the
strip width, at intermediate values of $\beta/\alpha$ in the
interval $[0,1]$. This behavior of the spin Hall conductivity was
correlated with the one for the longitudinal spin conductivity. This
study was then extended to the out-of-equilibrium regime that
arises by imposing a finite voltage bias between the two ends of an
open strip. This system, in the presence of a Hubbard term with
coupling $U$,
was treated with the density matrix renormalization group technique
and with the Landauer-Buttiker formalism. It was found that relevant
properties to the spin Hall effect, such as the transversal spin
current and the spin accumulation, present a similar nonmonotonic
behavior as the one found for the spin conductivities. More
importantly, it was also found that these properties are enhanced
by the repulsive Hubbard interaction up to a moderate value of $U$.
\end{abstract}

\maketitle

\section{Introduction}
\label{introsect}

The flow of spins in solids has recently received an intense interest
both because it manifests at a fundamental level in the field of
topological insulators \cite{ando-topins} and also because it may lead
to technological applications in spintronics \cite{prinz,wolf,zutic}.
Particularly important is the case when the spin flow is generated by
itinerant spin-orbit interactions (SOI) of the Rashba or Dresselhaus
forms.

In bulk inversion asymmetric (BIA) systems the SOI effectively leads
to the (linear) Dresselhaus spin-orbit coupling (DSOC)
\cite{dresselhaus}:
\begin{eqnarray}
H_{BIA} = \beta (\sigma_x k_x - \sigma_y k_y)
\end{eqnarray}
In most materials, this linear in momentum DSOC is accompanied by a 
term which is cubic in momentum but it will not be included in the
present study.
On the other hand, the structure inversion asymmetry (SIA), which is
due to the presence of surfaces or interfaces, the SOI gives rise
at an effective level to the Rashba spin-orbit coupling (RSOC) 
\cite{rashba,manchonrev2015} defined by the Hamiltonian:
\begin{eqnarray}
H_{SIA} = \alpha (\sigma_x k_y - \sigma_y k_x) =
\alpha {\boldsymbol \sigma}
\cdot [{\bf k} \times {\bf z}]
\end{eqnarray}
The RSOC has the important property of being able to be tuned by an
external electric field in addition to its value determined by the
intrinsic properties of the system.

In most materials, both Rashba and Dresselhaus SOI are present, and
their relative strength can be determined using magneto-transport
properties \cite{ganichev,ganichev16}, particularly by measuring the
beating patterns in Shubnikov-de Haas oscillations \cite{herzog}.
The presence of both types of SOI in a given system depends on the
atomic structure of the material involved but also on the direction on
which the wires are grown \cite{rainis}. The electric fields implied
in the lateral confinement that is frequently used to define
a strip can also modify the ratio between the RSOC and DSOC strengths
\cite{rainis,anghel}.

Both Rashba and Dresselhaus SOI lead to the spin Hall effect
but their effects interfere and interesting physics appears when
the RSOC and DSOC strengths, $\alpha$ and $\beta$, are varied.
The most interesting magnetic state is the so-called persistent
spin helix (PSH) \cite{bernevig}, which appears when both Rashba
and Dresselhaus are present with equal strength. In addition, 
at the PSH point, D'yakonov-Perel and Elliott-Yafet spin-flip
processes due to non-magnetic impurities are suppressed, thus
enabling non-ballistic transport \cite{wenk,dettwiler,schliemann}.
A clear signature of the persistent helix state was detected
as a dip in the magnetoconductance \cite{kohda}.

For arbitrary values of the ratio $\beta/\alpha$, the interband
contribution to the longitudinal optical conductivity was
examined for the isotropic two-dimensional (2D) system with a
parabolic band in the clean limit \cite{carbotte}, and it was found
that it disappears when $\alpha=\beta$. The spin Hall conductivity
was originally computed for the pure Rashba model on the infinite
plane for a parabolic band \cite{sinova04,rashba04}. This quantity
was also computed for arbitrary values of $\alpha,\beta$, and it
was found to be $\sigma_{sH}=\frac{1}{8\pi ^{2}}\gamma_{\pm }$ (in
our units of $e=1$) \cite{shen2004}, where $\gamma_{\pm }=
sign(\alpha^{2}-\beta^{2})$ is the Berry phase. That is,
$\sigma_{sH}=\frac{1}{8\pi}$ in the interval 
$0 \le \beta/\alpha < 1$.

Although most of the previous theoretical work has been done in the
isotropic 2D system, actual spintronics devices involve finite width
conductors or wires, and taking into account the nanoelectronics drive
towards smaller wire widths, it is of fundamental importance to study
the behavior of the relevant magneto-transport properties for the
smallest possible widths.
The relatively few studies on anisotropic 2D systems
were performed using electrostatic lateral confinement. In those
works various finite size effects were analyzed both theoretically
\cite{wenk,chang} and experimentally \cite{kohda,altmann,sasaki}.

The final ingredients for the model to be studied in the present work
come from emergent phenomena at oxide interfaces, particularly 
LaAlO$_3$/SrTiO$_3$ where RSOC is present \cite{hwang,caviglia,
banerjee,gopinadhan,caprara,khalsa,bucheli,ruhman}. While for
conventional semiconductor materials where spin-orbit effects were
studied, typically small electron fillings were considered, these new
materials motivate the research at larger electron fillings, where
electron correlations become more relevant and various magnetic
orderings induce complex transport behaviors.

The simplest and perhaps more interesting way of considering
electron-electron correlations is adding to the tight-binding
Hamiltonian the on-site Hubbard term. There are various studies of
models including SO and Hubbard interactions in one- and two-
dimensions. Most of these studies have only considered the Rashba
SOC \cite{riera13,goth,kocharian,hamad} but there are also studies
where both RSOC and DSOC were involved \cite{fsun2017}.

In the present work, various magneto-transport properties,
particularly those related to the spin-Hall effect, will be studied
on finite strips with Rashba and Dresselhaus SOI, in the presence
of electron-electron Hubbard interaction, in the whole range of
parameters $0 \le \beta/\alpha \le 1$, and for electron filling
$n=0.5$. These strips have
atomically defined open edges, that is, they are not regions
of a 2D system laterally delimited by voltage gates. Hence, the 
considered values of the Rashba and Dresselhaus SOI are intrinsic to
the material of the strip. This study will only  consider clean
systems.

The study of this model in out of equilibrium and interacting regimes
will be performed using computational techniques, the density matrix
renormalization group technique, and the Landauer-Buttiker formalism.
Although there are previous studies of 
channels  at particular values of the Rashba and Dresselhaus SO
couplings as a function of the applied voltage bias by the Landauer
formalism \cite{mhliu}, a systematic study as a function of these
couplings, as well as including electron
correlation effects, is still lacking.

This paper is organized as follows. In Sec.~\ref{modelsect} 
the second quantized model to be studied is defined, and the main
theoretical methods employed are outlined. In Sec.~\ref{noninter}
the study of the equilibrium non-interacting case is studied in
linear response. In Sec.~\ref{outofeq} the out of equilibrium
interacting two-chain strip is studied using density matrix 
renormalization group. In Sec.~\ref{landabut} the out of 
equilibrium system is studied within the Landauer-Buttiker
formalism. Finally, in the Conclusions, the main results obtained
are emphasized and their possible relevance to spintronics devices
is discussed.

\section{Model and methods}
\label{modelsect}

The Hamiltonian to be studied in the present work can be expressed
as $H = H_h + H_R + H_D + H_U $, where $H_h$ corresponds to the usual
hopping term:
\begin{align}
H_h = - t \sum_{<l,m>,\sigma} (c_{l,\sigma}^\dagger c_{m,\sigma} +
H. c.)
\label{hamhop}
\end{align}
The Rashba SO Hamiltonian on the square lattice in the $\{x,y\}$-plane
is given by \cite{pareek,XiaoChen,riera13,sinova14}:
\begin{align}
H_R = \alpha \sum_{l} &[c_{l+x,\downarrow}^\dagger c_{l,\uparrow} -
c_{l+x,\uparrow}^\dagger c_{l,\downarrow} + i (
c_{l+y,\downarrow}^\dagger c_{l,\uparrow} \nonumber  \\
&+ c_{l+y,\uparrow}^\dagger c_{l,\downarrow}) + H. c.]
\label{Rham}
\end{align}
and the Dresselhaus SO term is similarly given by (Appendix
\ref{app:RDham}):
\begin{align}
H_D = \beta \sum_{l} &[c_{l+y,\downarrow}^\dagger c_{l,\uparrow} -
c_{l+y,\uparrow}^\dagger c_{l,\downarrow} + i (
c_{l+x,\downarrow}^\dagger c_{l,\uparrow} \nonumber  \\
&+ c_{l+x,\uparrow}^\dagger c_{l,\downarrow}) + H. c.]
\label{Dham}
\end{align}
The last term of $H$ corresponds to the Hubbard interaction:
\begin{align}
H_U = U \sum_{l} n_{l,\uparrow} n_{l,\downarrow}
\label{hamhub}
\end{align}
where the notation is conventional.

The total Hamiltonian $H$ will be studied on strips of length $L$ in
the longitudinal $x$-direction and width $W$ in the transversal
$y$-direction, with $W<L$. Open boundary conditions (BC) are imposed
on the transversal direction.

In the following, the normalizations $\sqrt{\alpha^2+\beta^2}=V_{SO}$
and $\sqrt{t^2+V_{SO}^2}=1$, which will be the scale of energy,
have been adopted. These two normalizations are essential to compare
quantities for different ratios of $\alpha$ and $\beta$ for a fixed
value of $V_{SO}/t$, as is the purpose of the present study. For
example, with these normalizations, the relative difference in total
energy for $W=8$, $V_{SO}/t=0.5$, in the whole range of
$0 \le \beta/\alpha \le 1$, with $U=0$, is less than 0.023 \%, much 
smaller than the relative differences obtained for the physical
quantities studied. In the same way, the total energy is approximately
constant as $V_{SO}/t$ is varied for a fixed value of $\beta/\alpha$.

In the noninteracting case ($U=0$) the main quantity that will be studied
is the spin-Hall conductivity, $\sigma_{sH}$. At equilibrium, in linear
response, $\sigma_{sH}$ is defined as the zero frequency limit of the
spin-charge transversal response function at
zero temperature \cite{rashba04,sinova04}:
\begin{align}
\sigma^{sc}_{xy}(\omega) &= -i \frac{1}{\pi N} \sum_{n}
      \sum_{m} 
      \frac{\langle \Psi_n | \hat{j}^s_y | \Psi_m \rangle
     \langle \Psi_m | \hat{j}_x | \Psi_n \rangle}{[(E_n-E_m)^2-\omega^2]}
\label{sHcond}
\end{align}
where $\hat{j}_x$ is the longitudinal charge current operator
and $\hat{j}^s_y$ is the transversal spin current operator. The charge
current operator can be written as the sum of two
terms, $\hat{j}_{hop,x}$ and $\hat{j}_{SO,x}$, usually referred to as
the spin-conserving and spin-flipping currents, respectively
(see Appendix \ref{app:curop}). Similarly, the spin 
currents can be written as
$\hat{j}^s_y=\hat{j}^s_{hop,y} + \hat{j}^s_{SO,y}$.

In linear response, a measure of the spin current in the longitudinal
direction is given by the longitudinal spin conductivity 
$\sigma^{s}_{xx}$, which is defined as the zero frequency limit of
a response function analogous to the one given by Eq.~(\ref{sHcond})
except that the operator of the transversal spin current $\hat{j}^s_y$
is replaced by the operator of the longitudinal spin current
$\hat{j}^s_x$. Similarly, the anomalous Hall conductivity
$\sigma_{AH}$ corresponds to the zero frequency limit of a response
function obtained from Eq.~(\ref{sHcond}) by replacing $\hat{j}^s_y$
by the transversal charge current operator $\hat{j}_y$.

Of particular interest is the contribution from the spin-flipping
currents to the Drude peak, which will also be computed in linear
response. The hopping and SO contributions to the Drude weight are
defined as
\cite{riera17}:
\begin{align}
\frac{D_{a}}{2\pi} = \frac{K_{a,x}}{2N} - I_{reg,a}
\label{drude}
\end{align}
$a=hop, SO$, where $K_{hop,x}=-\langle (H_h)_x\rangle$,
$K_{SO,x}=-\langle (H_R+H_D)_x\rangle$, and
\begin{align}
I_{reg,a} = \frac{1}{N} \sum_{n \neq 0}
\frac{|\langle \Psi_n |\hat{j}_{a,x} |\Psi_0 \rangle |^2}{E_n-E_0}
\label{intreg}
\end{align}
are the corresponding contributions to the regular part of the
longitudinal optical conductivity. Notice that the total $I_{reg}$,
and hence the total Drude weight, is the sum of the hopping and SO
contributions, and the contribution that results from the product of
the matrix elements of $\hat{j}_{SO,x}$ and $\hat{j}_{hop,x}$. For all
the parameters considered in the present effort, this mixing
contribution is negligible \cite{note1}.

Linear response results were
obtained for strips with periodic BC along the longitudinal direction
by exact numerical diagonalization of the Hamiltonian in
momentum space (Appendix \ref{app:RDham}).

The study of out-of-equilibrium regimes, and in the presence of the
Hubbard interaction, is performed, for $W=2$ strips, by using the 
time-dependent density matrix-renormalization group (td-DMRG)
method \cite{schollwock,schmitteckert,alhassanieh}. In this case,
a small finite voltage bias, $\Delta V$, is imposed at the strip
ends at time $\tau=0$, after the ground state has been properly
described at equilibrium. This setup is schematically shown in
Fig.~\ref{fig4}(a) below.
This technique has been already employed to study two- and
three-chains Rashba-Hubbard strips \cite{riera13,hamad}.

In general, the time evolution of an arbitrary operator $\hat{O}$, 
is given by $O(\tau)=\langle \Psi(\tau)|\hat{O}|\Psi(\tau) \rangle$,
where $|\Psi(\tau) \rangle$ is the time-evolved ground state. The
procedure follows the so-called "static" algorithm \cite{schollwock}
and at each time step the time evolution operator is applied as a
series expansion involving up to the 40-th order. Then, the
time-evolution of several physical properties, such as charge and
spin currents, can be computed. The hopping contribution to the 
longitudinal charge current is directly computed from the operator
$\hat{j}_{hop,x}$, and the total longitudinal current is computed
as the time derivative of the charge occupation of one half of the
strip.

Although some qualitative features can be inferred by simple
inspection of $O(\tau)$ plots, for a more quantitative statement it
is necessary to assign a single number to each physical property
for any set of parameters $\beta/\alpha$, $V_{SO}/t$, and $W$.
Following Refs.~\cite{schollwock,schmitteckert,alhassanieh} this
single number corresponds to the amplitude of the time oscillation
that follow each physical property due to the finite length of the
system. That is, $O=amplitude(O(\tau))$. Further details will be
provided in Sec.~\ref{outofeq}.

In addition to transversal spin currents, another quantity related to
the spin Hall effect is the spin accumulation defined as:
\begin{align}
\Delta S^z = S^z_{e1}- S^z_{e2}
\label{spinacc}
\end{align}
where:
\begin{align}
S^z_{e1}&=\sum_{i=1,w} S^z(i)  \nonumber\\
S^z_{e2}&=\sum_{i=1,w} S^z(W-w+1)
\nonumber
\end{align}
where $S^z(i)$ is the 
total $z$-magnetization of leg $i$ ($i=1,\ldots,W$), and $w=max(1,W/4)$.

An alternative approach, suitable to study wider strips, is the
Landauer-Buttiker approach, which deals with the transmission of an
electron wavepacket through a finite ''scattering region", described
by the total Hamiltonian $H$ ($U=0$) connected to two semi-infinite
leads (horizontal leads in Fig.~\ref{fig7}(a)) described by $H_h$. A 
small voltage bias $\Delta V=0.1$ is applied between these two
semi-infinite leads. Calculations within this approach are performed
using the Kwant package \cite{kwant} at zero temperature and at
quarter filling, which is imposed by appropriately setting the Fermi
level of the central, scattering region \cite{note}. Rather than
computing conductances, a procedure giving microscopic quantities
such as the $x,y,z$-components of the spin at each site and the charge
and spin currents between each nearest neighbor pair of sites, was
adopted \cite{nikolic,nikolic2014}. 
These quantities are obtained by taking the quantum averages of the
corresponding operators over the scattering wave function on the
scattering region. In order to compute the spin accumulation and
the transversal spin currents averaged over a region at the strip
edge, a four-terminal setup is used (Fig.~\ref{fig7}(a)). The 
vertical semiinfinite leads are also described by $H_h$. The code
has been checked by verifying that the spin currents and
$<S^z_l>$, $l=1 \ldots N$ (and hence the spin accumulation) vanish at
the $\beta=\alpha$ point, and also at this point $<S^x_l>=<S^y_l>$ due
to the restoring of the $U(1)$ symmetry. In addition, by 
interchanging $\alpha$ and $\beta$, the same results are obtained
with reversed sign for the transversal spin currents and $<S^z_l>$.

For the interacting case, $U >0$, a simple Hartree-Fock decoupling
is implemented and at each site the values of
$\langle n_\uparrow \rangle$, $\langle n_\downarrow \rangle$, are
provided by independent variational Monte Carlo simulations involving
a single Gutzwiller factor for the Hubbard repulsion $U$ 
\cite{giamarchi,hamad}. Additional details of the calculation of
these properties are provided in Sec.~\ref{landabut}. 

\begin{figure}[t]
\includegraphics[width=0.9\columnwidth,angle=0]{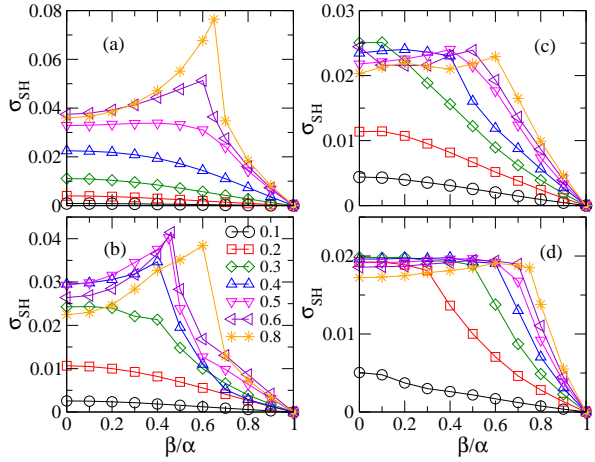}
\caption{(Color online) Spin Hall conductivity as a function
of $\beta/\alpha$, for various $V_{SO}/t$ indicated on the plot, (a)
$W=2$, (b) $W=4$, (c) $W=8$, and (d) $W=32$.}
\label{fig1}
\end{figure}

\section{Noninteracting strips, linear response}
\label{noninter}

Let us start with the noninteracting case, $U=0$, at equilibrium. All
the results shown in the present section were obtained for strips of
length $L = 2000$ with periodic BC in the longitudinal direction.

Results for the spin Hall conductivity $\sigma_{sH}$, obtained using
Eq.~(\ref{sHcond}), as a function of $\beta/\alpha$ for various strip
widths $W$ and SOI strengths $V_{SO}/t$ are shown in Fig.~\ref{fig1}.
For all the strip widths and $V_{SO}/t$ considered, $\sigma_{sH}$
presents a finite value at the Rashba point ($\beta=0$) and vanishes
at the PSH point ($\beta=\alpha$). This latter result is expected
because $\sigma_{sH}$ reverses it sign when the values of $\beta$ and
$\alpha$ are interchanged. For small values of $V_{SO}/t$ and small
$W$, the expected monotonic decrease of $\sigma_{sH}$ as
$\beta/\alpha$ varies from zero to one is observed.

However, as it can be seen in Fig.~\ref{fig1}(a) for $W = 2$, there
is an unexpected nonmonotonic behavior as $\beta/\alpha$ increases
from zero to one for large values of $V_{SO}/t$. This nonmonotonic
behavior is one of the main results of this work. For $W = 2$ it is
signalled by the presence of a peak in $\sigma_{sH}$ at an
intermediate value of
$\beta/\alpha$, $R^*\approx 0.6$, for $V_{SO}/t \gtrapprox 0.6$. For
$W = 4$, (Fig.~\ref{fig1}(b)) this peak is already present for
$V_{SO}/t \ge 0.4$, although a small cusp can be observed for
$V_{SO}/t = 0.3$. The position of the peak $R^*$ moves from
$\approx 0.4$ to $\approx 0.6$ as $V_{SO}/t$ increases. 

The peak in $\sigma_{sH}$ at $R^*$, for the same value of $V_{SO}/t$,
is most intense for $W = 2$, and it becomes less pronounced as
the strip width is increased. Although the peak is still present for
$W = 8$ (Fig.~\ref{fig1}(c)), it has mostly disappeared and changed
into a cusp for $W = 32$
(Fig.~\ref{fig1}(d)). For $W = 64$, results are virtually
indistinguishable from those of $W = 32$. For $W \ge 32$,
$\sigma_{sH}$ becomes approximately constant in the interval
$0 \le \beta/\alpha \le 1-\epsilon$, with 
$\epsilon \rightarrow 0$ for $V_{SO}/t \rightarrow \infty$.

\begin{figure}[t]
\includegraphics[width=0.9\columnwidth,angle=0]{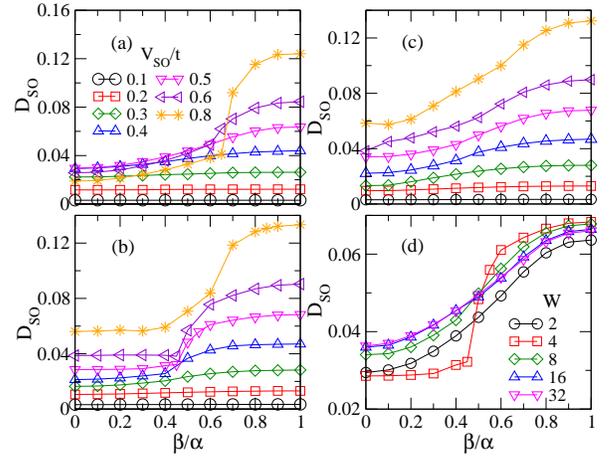}
\caption{(Color online) Spin-flipping contribution to the Drude peak,
$D_{SO}$, as a function of $\beta/\alpha$, for various $V_{SO}/t$
indicated on the plot, (a) $W=2$, (b) $W=4$, (c) $W=8$.
(d) $D_{SO}$ for $V_{SO}/t=0.5$, and various strip widths $W$
indicated on the plot.}
\label{fig2}
\end{figure}

It is also important to remark that this peak or cusp in $\sigma_{sH}$
separates two clearly different regimes with different curvatures for
$\beta/\alpha$ below or above its position $R^*$.

By replacing the two contributions to both charge and spin currents,
as discussed in the previous section,  into the integrand of
Eq.~(\ref{sHcond}), it turns out that there are four possible
contributions to $\sigma_{sH}$. For strips with periodic BC on the
longitudinal $x$ direction, only the contribution from the product of
the matrix elements of $\hat{j}_{SO,x}$ and $\hat{j}^s_{hop,y}$ is
different from zero for all values of $\beta/\alpha$, $V_{SO}/t$ and
$W$ considered, thus extending the previous result for the
pure Rashba model \cite{riera17}. This behavior also holds when open
BC are adopted in the longitudinal direction for sufficiently long
chains, but as the length $L$ is reduced, other contributions
become sizable particularly the one involving the product of the
matrix elements of $\hat{j}_{hop,x}$ and
$\hat{j}^s_{hop,y}$.

Since the spin-flipping part of the longitudinal charge current
is correlated with the hopping part of the transversal spin to produce
a finite value of the spin Hall conductivity, it is interesting to
examine how $\hat{j}_{SO,x}$ correlates with the operators involved
in other physical properties as a function of $\beta/\alpha$.

Let us study in the first place the contribution of $\hat{j}_{SO,x}$
to the Drude peak, as defined in Sec.~\ref{modelsect}. Results for
various strip widths and SOI strengths $V_{SO}/t$ are shown in
Fig.~\ref{fig2}. A general trend of increasing $D_{SO}$ with
$\beta/\alpha$ for a given $V_{SO}/t$ and $W$ can be observed. For a 
fixed $W$ and $\beta/\alpha$, there is also a general increase of
$D_{SO}$ with $V_{SO}/t$, as expected, with the exception of $W=2$
and $V_{SO}/t \gtrapprox 0.6$, for small values of $\beta/\alpha$, as
it can be observed in Fig.~\ref{fig2}(a). More relevant for the
discussion of the spin Hall conductivity is the presence of a jump in
$D_{SO}$ at the value of $\beta/\alpha =R^*$ at which there is a peak
in $\sigma_{sH}$ for the corresponding values of $V_{SO}/t$, for
$W=2$ and 4, as it can be seen in Figs.~\ref{fig2}(a) and
\ref{fig2}(b). In contrast, for $W=8$, consistently with the
smoothing out of the peaks in $\sigma_{sH}$, the jumps are replaced
by an inflection point for the
corresponding values of $V_{SO}/t$, as shown in
Fig.~\ref{fig2}(c). The dependence of $D_{SO}$ with $\beta/\alpha$
for $W \gtrapprox 16$ becomes increasingly smooth as $W$ is increased.
In order to
make this behavior more apparent, $D_{SO}$ was replotted in
Fig.~\ref{fig2}(d) for a single value of $V_{SO}/t=0.5$, and various
values of $W$. It is also noticeable that the dependence with $W$ is
already saturated for $W \approx 32$.

In the second place, let us examine the contribution of
$\hat{j}_{SO,x}$ to another magneto-transport effect
due to the SOI that is the longitudinal 
spin conductivity $\sigma^{s}_{xx}$, which is the linear response
corresponding to the spin polarized current to be studied in the
next sections.

\begin{figure}[t]
\includegraphics[width=0.9\columnwidth,angle=0]{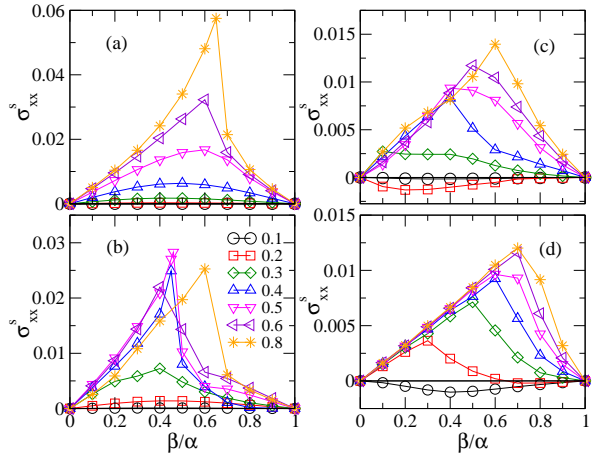}
\caption{(Color online) Longitudinal spin conductivity as a function
of $\beta/\alpha$, for various $V_{SO}/t$ indicated on the plot, (a)
$W=2$, (b) $W=4$, (c) $W=8$, and (d) $W=32$.}
\label{fig3}
\end{figure}

\begin{figure}[t]
\includegraphics[width=0.95\columnwidth,angle=0]{dmrgtime.eps}
\put(-220,98){
\includegraphics[width=0.37\columnwidth,angle=0]{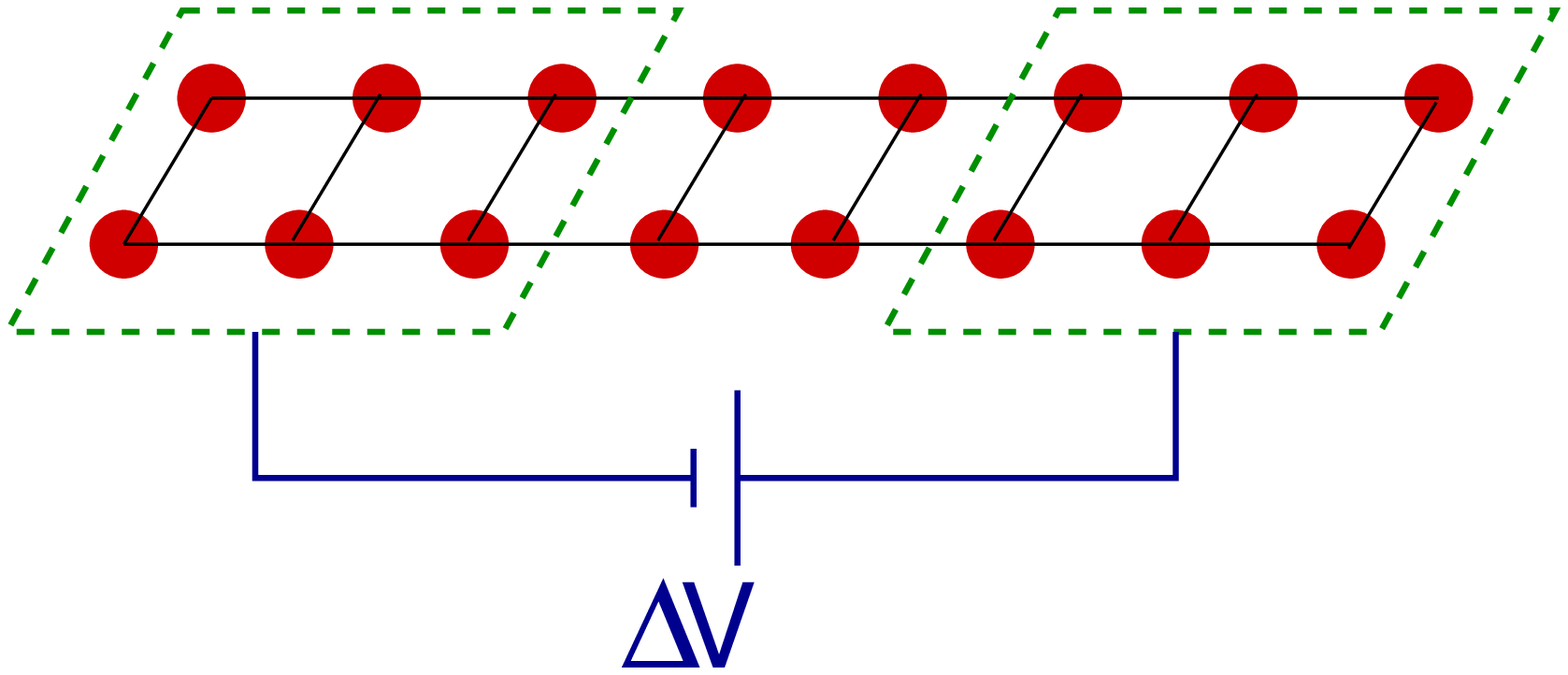}}
\caption{(Color online) (a) Schematic picture of the td-DMRG setup.
(b) Time evolution of the longitudinal SO or spin-flipping charge
current, $J_{SO,x}$, for $U=0$, 2, 4 and 6, $\beta=0$ (full lines)
and $\beta=\alpha$ (dashed lines), $V_{SO}/t=0.6$. In (b), results
for $U=0$ and $\beta/\alpha=0.2$, 0.3, 0.4, 0.5, 0.6, and 0.7
(from bottom to top) are shown with grey lines.
(c) Amplitude of $J_{SO,x}$ (see text), as a function of
$\beta/\alpha$, for $V_{SO}/t=0.2$ (circles), 0.4 (squares), 0.5
(up triangles), 0.6 (diamonds), and 0.8 (down triangles), $U=0$.
(d) Amplitude of $J_{SO,x}$ as a function of $\beta/\alpha$,
for $V_{SO}/t=0.6$, and $U=0$, 1, 2, 4, 6, and 8, from top to
bottom. Results obtained with td-DMRG on the $24 \times 2$ cluster.
}
\label{fig4}
\end{figure}

Results for the longitudinal spin conductivity for various values of
$W$ and $V_{SO}/t$ as a function of $\beta/\alpha$ are shown in
Fig.~\ref{fig3}.  First, notice that this quantity vanishes at the
pure Rashba and at the pure Dresselhaus points, which is a well-known
result \cite{inoue2003,amin2016}. As expected for the same argument
as for the spin Hall conductivity, $\sigma^{s}_{xx}$ also vanishes
at $\beta=\alpha$. For $W=2$
(Fig.~\ref{fig3}(a)), $\sigma^{s}_{xx}$ acquire finite values by
increasing $\beta/\alpha$ reaching a maximum at an intermediate point.
This dependence is smooth for small values of $V_{SO}/t$. For 
$V_{SO}/t \ge 0.6$, a sharp peak appears separating two regions with
quite different slopes. A similar behavior can be observed for 
$W=4$ (Fig.~\ref{fig3}(b)), For $W=2$ and 4, the smooth behavior
of $\sigma^{s}_{xx}$ occurs for the same values of $V_{SO}/t$ for
which a smooth behavior is present in $\sigma_{sH}$,
and for large $V_{SO}/t$, the cusps occur also at the same values
$R^*$ as the cusps in $\sigma_{sH}$ for the same $V_{SO}/t$, as seen
in Figs.~\ref{fig1}(a) and \ref{fig1}(b).

This correspondence between the behaviors of $\sigma_{sH}$ and
$\sigma^{s}_{xx}$, for the same values of $V_{SO}/t$ and $W$, 
suggests that the latter could be obtained from the former by
subtracting a quantity that decreases linearly from its maximum value
at $\beta =0$ to zero at the PSH point. For large strip widths, as
shown in Figs.~\ref{fig3}(c) and \ref{fig3}(d) for $W=8$ and 32
respectively, $\sigma^{s}_{xx}$ presents a well defined linear
behavior for small $\beta/\alpha$ that extends up to a value
$1-\epsilon$, where $\epsilon$ is equal to the value described above
for $\sigma_{sH}$ for the same $V_{SO}/t$ and $W$. Hence, these
behaviors for large $W$ and $V_{SO}/t$ gives further support to the
previous suggestion that $\sigma^{s}_{xx}$ and $\sigma_{sH}$ differ
by a linear function decreasing from $\beta=0$
to $\beta=\alpha$.

Let us now come back to the previous discussion about the role of
$\hat{j}_{SO,x}$. As for the $\sigma_{sH}$ case, the integrand in
$\sigma^{s}_{xx}$ can be split in four contributions, and again the
solely nonvanishing contribution turns out to be the one involving
the matrix elements of $\hat{j}_{SO,x}$, this time multiplied by the
matrix elements of $\hat{j}^s_{hop,x}$. This fact emphasizes the
central role played by the longitudinal spin flipping current
$\hat{j}_{SO,x}$ in the two most relevant effects of itinerant SOI.

Just for completeness, and partially for checking purposes, the
anomalous Hall conductivity, extending the well-known result
for the pure Rashba model \cite{AHERMP}, was found to vanish
for all values of $V_{SO}/t$, $W$, and $\beta/\alpha$ considered.

\section{Out of equilibrium regime}
\label{outofeq}

As mentioned in Sec.~\ref{modelsect}, it is necessary to resort 
to techniques such as td-DMRG to study properties in out of
equilibrium regimes and in the presence of electron-electron
interactions. In the present section, td-DMRG is applied only 
to the $W=2$ strip, specifically to the $24\times 2$ system at
$n=0.5$. Most results were obtained by retaining 600-700 states in
the truncation stage. A schematic illustration of the
computational setup is shown in Fig.~\ref{fig4}(a). The voltage
bias $\Delta V=0.01$ is applied at time $\tau=0$ to the two halves
of the system.

Typical td-DMRG time-evolution results for the SO contribution to
the longitudinal charge current, $J_{SO,x}$, are provided in
Fig.~\ref{fig4}(b). This quantity was selected because it
corresponds to the operator $\hat{j}_{SO,x}$, which plays an
essential role in the behavior of relevant properties in
linear response, as discussed in the previous section. In addition,
between the components of the total longitudinal current $J_x$,
$J_{SO,x}$ is the one that has the strongest dependence
with $\beta/\alpha$. as it can be observed in Fig.~\ref{fig4}(b)
for $V_{SO}/t=0.6$, and for various values of $U$.

As described in Sec.~\ref{modelsect}, the value of each quantity
for each set of parameters is adopted as the amplitude of its time
evolution. As it can be seen in Fig.~\ref{fig4}(b), for most 
properties, this time evolution presents at small times a double peak
structure, although in some cases one of the peaks appears as a
shoulder. Due to the relatively few states retained, and the 
algorithm adopted, only the results at short times are reliable.
Then, by convention, the amplitude is defined as the average
of the time evolution between those first two peaks.

Results for $J_{SO,x}$ as a function of $\beta/\alpha$ for various
values of $V_{SO}/t$ and $U=0$ are shown in Fig.~\ref{fig4}(c). In
the first place, $J_{SO,x}$ increases in general, as expected, with
$V_{SO}/t$, for all the interval of $\beta/\alpha$ considered except
near the pure Rashba model for large $V_{SO}/t$, as it was reported
before in Ref.~\cite{riera13}. More important is
that for a fixed $V_{SO}/t$, $J_{SO,x}$ increases with
$\beta/\alpha$ and this behavior becomes most pronounced as
$V_{SO}/t$ increases. Overall, the behavior of $J_{SO,x}$ with
$\beta/\alpha$ and $V_{SO}/t$ follows very closely the one for 
$D_{SO}$ shown in Fig.~\ref{fig2}(a).

The effect of the Hubbard repulsion is in general, as it is well-known
in correlated electron metallic systems, to suppress charge currents.
This effect is already apparent in Fig.~\ref{fig4}(b), for both the
total current $J_x$ and its $J_{SO,x}$ contribution. A more systematic
and quantitative study of the variation of $J_{SO,x}$ with $U$ as a
function of $\beta/\alpha$ and for $V_{SO}/t=0.6$, is provided in
Fig.~\ref{fig4}(d). Indeed, as it can be seen in this Figure, the
suppression of $J_{SO,x}$ with $U$ takes place up to the largest
Hubbard repulsion considered, $U=8$, while the system remains in its
metallic state. The same behavior is observed for all values of
$V_{SO}/t$.

Since the transversal spin current $J^s_y$ involves differences
between various terms (see Appendix~\ref{app:curop}), in order to
avoid large errors stemming from the separate time evolution of each
of those terms, it is preferable to compute this current as a time
derivative of the total $S^z$ of the sites located on two rungs at  
the center of the strip and belonging to the same chain. In this way,
the total $J^s_y$ is computed but the
separate information on $J^s_{hop,y}$ and $J^s_{SO,y}$ is lost.
To examine the total $J^s_y$ instead of the more relevant, according
to the results obtained in linear response,
$J^s_{hop,y}$, is in any case innocuous since the contribution from
$J^s_{SO,y}$ is always much smaller, as it will be discussed
in Sec.~\ref{landabut}

\begin{figure}[t]
\includegraphics[width=0.9\columnwidth,angle=0]{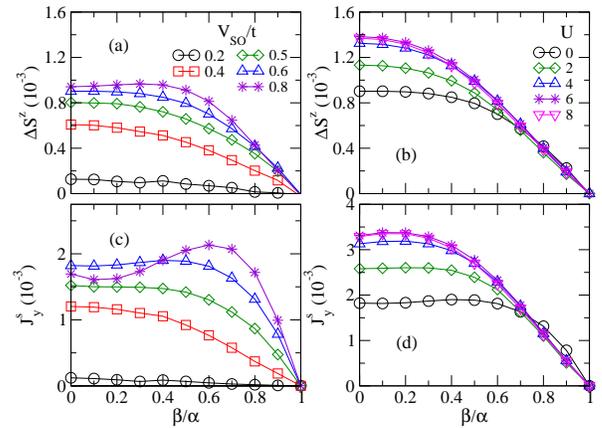}
\caption{(Color online) (a, b) Spin accumulation and (c, d)
$J^s_y$, as a function of $\beta/\alpha$. (a, c) Results
for various values of $V_{SO}/t$ (symbols indicated on the plot)
and $U=0$. (b, d) Results for $V_{SO}/t=0.6$ and various values
of $U$ indicated on the plot. Results obtained with td-DMRG
on the $24 \times 2$ cluster.}
\label{fig5}
\end{figure}

Results obtained by td-DMRG for the two most relevant properties
in the context of the spin Hall effect, the transversal spin current
$J^s_y$ and the spin accumulation $\Delta S^z$, defined by 
Eq.~(\ref{spinacc}), are shown in Fig.~\ref{fig5}.

Let us start with the noninteracting case, $U=0$. Results for
$\Delta S^z$ and $J^s_y$ as a function of $\beta/\alpha$, for various
$V_{SO}/t$, are shown in Figs.~\ref{fig5}(a) and \ref{fig5}(c)
respectively. In general, as expected, these quantities increase with
$V_{SO}/t$ for any value of $\beta/\alpha$. As for $J_{SO,x}$, a
departure of this behavior can be observed for $J^s_y$ near the
Rashba limit and large $V_{SO}/t$, and the same behavior occurs for
$\sigma_{sH}$ in linear response (Fig.~\ref{fig1}(a)).
Another general behavior is the
vanishing of $J^s_{y}$ and $\Delta S^z$ as $\beta/\alpha$ approaches
1, that is as the system approaches the persistent helix state. This
behavior is again expected since by switching $\beta$ and $\alpha$,
$J^s_{y}$ and $\Delta S^z$ reverse their signs.

More importantly, it should be noticed that a nonmonotonic behavior
in $J^s_y$ can be observed for $V_{SO}/t \gtrapprox 0.6$ in
Fig.~\ref{fig5}(c). This
maximum is larger than error bars, which are of the order of the
symbol sizes. The presence of this maximum, although much less
pronounced, is consistent with the one reported previously for the
spin Hall conductivity (Fig.~\ref{fig1}). Taking into account
the opposite trends in the variation with $\beta/\alpha$ of
$J_{SO,x}$, shown in Fig.~\ref{fig4}(c) and of $J^s_y$
(Fig.~\ref{fig5}(b)), and since $\sigma_{sH}$ involves the commutator
between both quantities, one could speculate that a convolution of
those behaviors would lead to the strong peak observed in
$\sigma_{sH}$ at an intermediate value of $\beta/\alpha$.

On the other hand, the behavior of $\Delta S^z$ is monotonic between
$\beta=0$ and $\beta=\alpha$. The very weak maximum observed for
$V_{SO}/t=0.8$ certainly falls within the error bars
of the calculation.

Let us now examine the evolution of these quantities when the Hubbard
interaction $U$ is switched on. Figs.~\ref{fig5}(b) and \ref{fig5}(d)
show results for the spin accumulation and the transversal spin 
current, respectively, for  $V_{SO}/t=0.6$, and various values
of $U$. In Fig.~\ref{fig5}(b), the most noticeable feature is the
systematic increase of $\Delta S^z$ with $U$ up to the maximum value
considered, $U=8$, in the whole interval of $\beta/\alpha$, with the
constraint that  $\Delta S^z \rightarrow 0$ when
$\beta/\alpha \rightarrow 1$. Notice that results for $U=6$ and 
$U=8$ are indistinguishable.
This enhancement in $\Delta S^z$ was observed for all the values
of $V_{SO}/t$ examined, thus extending the result obtained
for the Rashba model \cite{riera13} to the whole range of
$\beta/\alpha$.

Similarly, as shown in Fig.~\ref{fig5}(d), $J^s_y$
also increases with $U$, again saturating at $U \approx 8$. Besides,
the presence of a maximum of $J^s_y$ at an
intermediate value of $\beta/\alpha$ is preserved by $U$. However,
this maximum is smoothed out by $U$ and its location is
shifted to lower values of $\beta/\alpha$.
This enhancement of $J^s_y$ with $U$ is observed for
$V_{SO}/t \leq 0.6$, but for $V_{SO}/t=0.8$, $J^s_y$ becomes actually
suppressed by increasing $U$. Notice that, as said above, even for
this value of $V_{SO}/t$, the spin accumulation is enhanced by $U$.
Since $J_{SO,x}$ is in general suppressed by $U$, while $J^s_y$ is
enhanced by $U$, the previously mentioned handwaving argument based
on a convolution of $J_{SO,x}$ and $J^s_y$, would not lead to a
conclusive guess for the behavior of the spin Hall conductivity
with $U$.

\begin{figure}[t]
\includegraphics[width=0.9\columnwidth,angle=0]{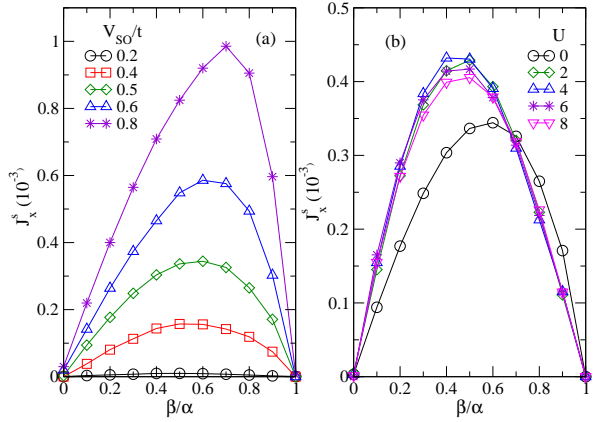}
\caption{(Color online) Longitudinal spin current $J^s_x$, as a
function of $\beta/\alpha$, (a) for various values of $V_{SO}/t$
(symbols indicated on the plot) and $U=0$, (b) for $V_{SO}/t=0.5$
and various values of $U$ indicated on the plot. Results obtained with
td-DMRG on the $24 \times 2$ cluster.}
\label{fig6}
\end{figure}

To end this section, let us examine the evolution of the longitudinal
or polarized spin current, $J^s_x$, as a function of $\beta/\alpha$.
The behavior of this quantity in this slightly out of equilibrium
system should be compared with the one of $\sigma^s_{xx}$ discussed
in the previous section. In the same way as for the
transversal spin current, in order to minimize errors, $J^s_x$ is
computed as the time derivative of the total $S^z$ of the left half
of the cluster. Again, the separate information of the hopping
or SO contributions to $J^s_x$ is lost but as discussed above, to
analyze $J^s_x$ instead of $J^s_{hop,x}$ is relatively innocuous,
and in any case, it is $J^s_x$ the quantity that is experimentally
accessible.

Results for $J^s_x$ as a function of $\beta/\alpha$ for various values
of $V_{SO}/t$, $U=0$, are shown in Fig.~\ref{fig6}(a), where it could
be observed that $J^s_x$ follows roughly the same behavior as 
$\sigma^s_{xx}$ shown in Fig.~\ref{fig3}(a). In particular, $J^s_x$
vanishes both at the Rashba point and at the PSH point.
The more rounded-off
dependence may be due to taking $J^s_x$ instead of $J^s_{hop,x}$, and
also due to the relatively small value of $L$ and open BC adopted.
In any case, it is clear that there is an asymmetric shape of $J^s_x$,
and that, for $V_{SO}/t=0.8$, there is a change of curvature at the
maximum value around $0.7 \le \beta/\alpha \le 0.8$.

In Fig.~\ref{fig6}(b), it can be seen that the longitudinal spin
current is first {\em enhanced} by the Hubbard repulsion, reaching a
maximum at $U=4$, and then it is {\em suppressed} for larger values
of $U$. That is, the behavior of $J^s_x$ with $U$ is similar to the
one of the spin accumulation and the transversal spin current, but
its dependence with $U$ is different. This issue would deserve 
further study, increasing the precision and examining finite size
effects, but this is out of reach of present computational 
capabilities,

\begin{figure}
\includegraphics[width=0.90\columnwidth,angle=0]{jx-kw-w4-8-16.eps}
\put(-210,97){
\includegraphics[width=0.38\columnwidth,angle=0]{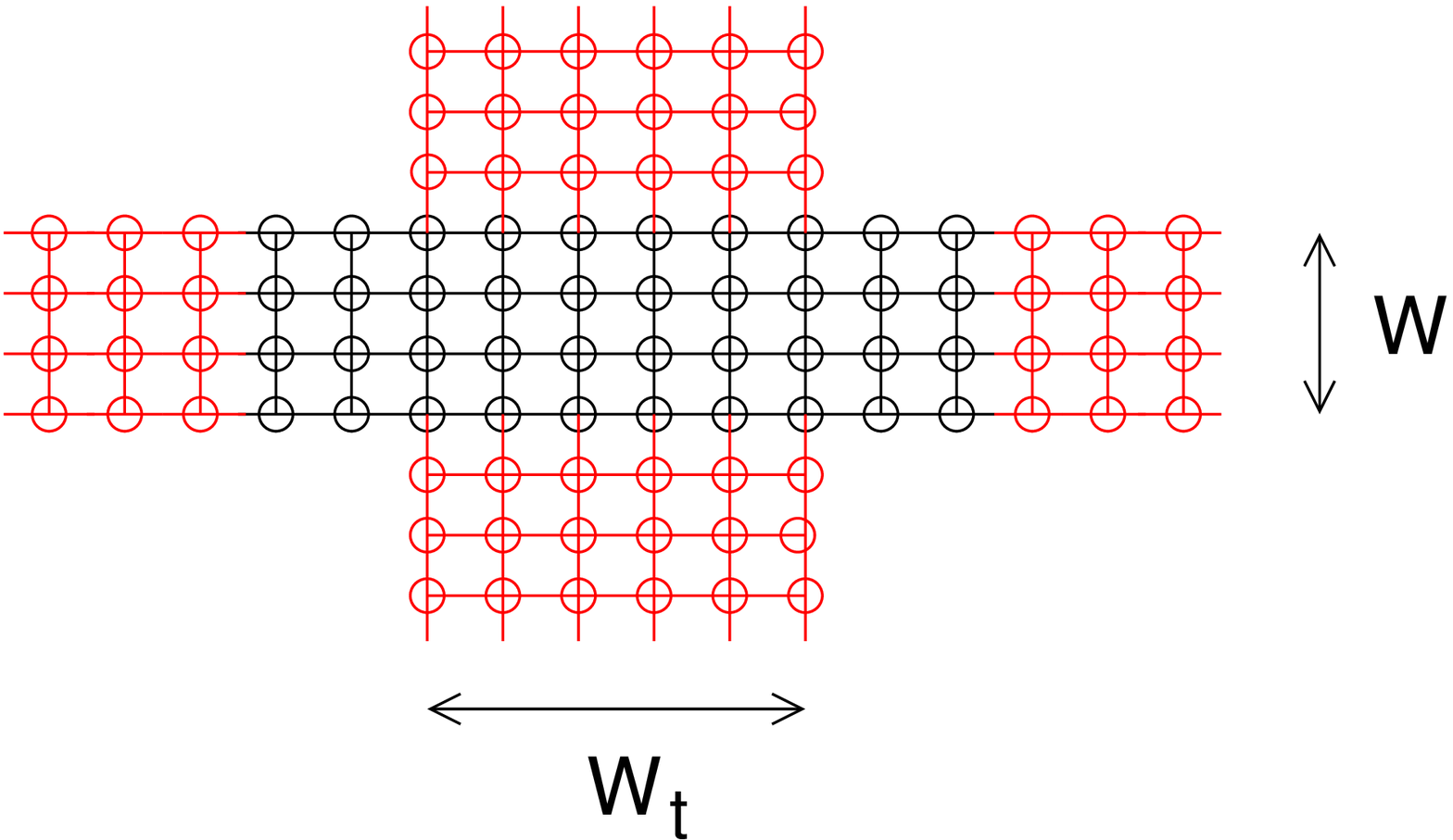}}
\caption{(Color online) (a) Schematic illustration of the four
terminal setup for the Landauer-Buttiker calculation. (b) $J_{SO,x}$
per chain as a function of $\beta/\alpha$ for $V_{SO}/t=0.2$, 0.4 0.5,
0.6 and 0.8, from bottom to top, obtained for the $80 \times 16$
cluster, $U=0$. (c) $J_{SO,x}/W$ for $V_{SO}/t=0.5$ and $U=0$,
obtained for the $40 \times 4$, $50 \times 8$, and $80 \times 16$
clusters from bottom to top. (d) $J_{SO,x}/W$ for $V_{SO}/t=0.5$ and
$U=0$, 1, 2, and 4, from top to bottom, obtained for the $50 \times 8$
cluster.}
\label{fig7}
\end{figure}

\section{Landauer-Buttiker approach}
\label{landabut}

The setup of the system is schematically shown in Fig.~\ref{fig7}(a).
The horizontal leads and the central scattering region have width $W$
and the two vertical transversal leads have width $W_t$. The four
leads are semi-infinite and the central region has length $L$. The
leads are numbered from 0 to 3 starting from the left horizontal
lead and moving clockwise. A small voltage bias 
$\Delta V_{02}=0.1$ is applied between the horizontal leads, and
$\Delta V_{01}=\Delta V_{03}=\Delta V_{02}/2$, which implies
$\Delta V_{13}=0$.

Fig.~\ref{fig7}(b) shows results for the SO or spin flipping
longitudinal charge current, $J_{SO,x}$, per chain, obtained on the
$80 \times 16$ central cluster for various values of $V_{SO}/t$, and
$U=0$. These results were obtained by averaging over $W_t=24$, 32 and
40. The error bars due solely to this averaging procedure are shown
for example for the $50 \times 8$ central region in
Fig.~\ref{fig7}(c). In general, $J_{SO,x}/W$ not only increases
with $V_{SO}/t$ for each $\beta/\alpha$ but, what is more important,
it increases with $\beta/\alpha$ for a fixed $V_{SO}/t$. This
behavior is more clear when a single value of $V_{SO}/t$ is
considered, as in Figs.~\ref{fig7}(c) and \ref{fig7}(d).

In Fig.~\ref{fig7}(c), $J_{SO,x}/W$ is shown as a function of
$\beta/\alpha$, for $V_{SO}/t=0.5$, and for central clusters of
varying width, $W=4$, 8, and 16. It can be observed not only an
overall increase of $J_{SO,x}/W$ with $W$, but also an increasing
slope. This overall increase with $W$ is consistent
with the one obtained in linear response for $D_{SO}$
(Fig.~\ref{fig2}(d)).  Results for the $40 \times 4$ and
$50 \times 8$ central clusters have been obtained by averaging over
$W_t=16$, 24 and 32.

In Fig.~\ref{fig7}(d), $J_{SO,x}/W$ is shown for $V_{SO}/t=0.5$, and
various values of $U$ on a $50 \times 8$ cluster. As expected, 
$J_{SO,x}/W$ decreases as $U$ is increased, as it was observed for
$W=2$ in Fig.~\ref{fig4}(d), but the system remains metallic up to
the largest value of the Hubbard repulsion considered, $U=4$. The
most relevant result is that the increasing trend of $J_{SO,x}/W$
with $\beta/\alpha$ is still clearly present up to $U=4$.

\begin{figure}[t]
\includegraphics[width=0.9\columnwidth,angle=0]{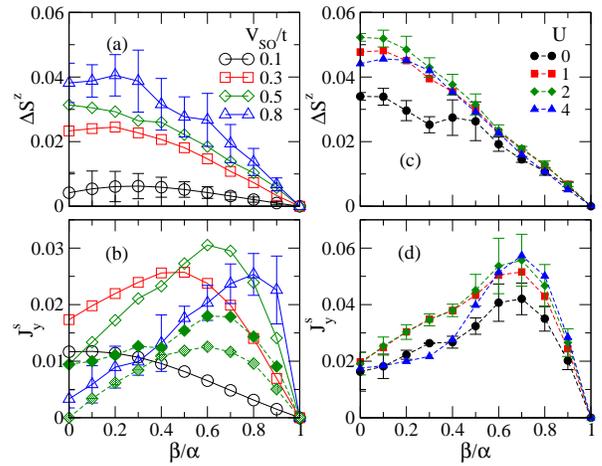}
\caption{(Color online) (a) Spin accumulation and (b) total
transversal spin current on the $200\times 8$ cluster, as a function
of $\beta/\alpha$ for various values of $V_{SO}/t$ as indicated on the
plot (open symbols, solid lines). (c) Spin accumulation and (d) total
transversal spin current for $V_{SO}/t=0.5$ for various values of $U$
indicated on the plot, on the $50\times 8$ cluster. In (b) the 
hopping (full diamonds, dashed line) and SO (shaded diamonds, dotted
line) contributions to $J^s_y$ for $V_{SO}/t=0.5$ have been added.
Results obtained by the Landauer-Buttiker formalism.}
\label{fig8}
\end{figure}

Let us now consider the two quantities that are more relevant to the
spin Hall effect, the spin accumulation and the transversal spin
current, starting with the noninteracting case, $U=0$. The central
system is a $200\times 8$ cluster. The error bars, shown in
Figs.~\ref{fig8}(a) and \ref{fig8}(b) for the two extreme values of
$V_{SO}/t$ considered, again only correspond to the averaging over
three different
widths of the transversal leads, $W_t=24$, 40 and 80. For these two
physical properties, the error bars are much larger than the ones for
$J_{SO,x}$.

Results for $\Delta S^z$ as a function of $\beta/\alpha$ for various
values of $V_{SO}/t$ are shown in Fig.~\ref{fig8}(a). By neglecting
some oscillations that are not significant within the errors of the
calculation, it can be seen that the main trends are that 
$\Delta S^z$ decreases monotonically for a given value of $V_{SO}/t$
as a function
of $\beta/\alpha$, vanishing at the PSH point, and increases at each
value of $\beta/\alpha$ by increasing $V_{SO}/t$. These behaviors are
similar to those for the $W=2$ strip obtained by td-DMRG and shown in
Fig.~\ref{fig5}(a).

More interesting are the results for the transversal spin
current, $J^s_y$, shown in Fig.~\ref{fig8}(b). In this 
case there is clear nonmonotonic behavior characterized by a maximum
of $J^s_y$ that is located at values of $\beta/\alpha$ that 
increase with $V_{SO}/t$. The presence of this maximum in $J^s_y$
confirms the behavior shown in Fig.~\ref{fig5}(c) for $W=2$, and is
consistent with the one for $\sigma_{sH}$ depicted in Fig.~\ref{fig1}.
Also notice that the hopping part of $J^s_y$ is in general larger
than the SO part, and these two contributions have the same sign, as
shown in Fig.~\ref{fig8}(b). In the Rashba limit, $\beta=0$,
$J^s_{SO,y}$ is strictly equal to zero for all $V_{SO}/t$, in
agreement with previous results \cite{riera17} and consistently with
the results for $\sigma_{sH}$ where the solely contributing matrix
elements are those of the operator $\hat{j}^s_{hop,x}$.

Let us now examine the effects of the Hubbard repulsion $U$.
The error bars of the Landauer-Buttiker part of the calculation
were obtained as before, but in this case to these errors one
should have to add the ones coming from the Hartree-Fock procedure,
The latter are difficult to estimate but they certainly increase with
$U$. Results for the spin accumulation $\Delta S^z$ on the
$50\times 8$ central scattering region for $V_{SO}/t=0.5$ are
depicted in Fig.~\ref{fig8}(c). The most remarkable behavior is the
enhancement of $\Delta S^z$ with $U$, in all the range of
$\beta/\alpha$, thus complementing the td-DMRG results for $W=2$ shown
in Fig.~\ref{fig5}(c). The maximum value of $\Delta S^z$ is reached
at $U=2$. Taking into account the likely error bars of the calculation
one could conclude that the behavior of $\Delta S^z$ as a function of
$\beta/\alpha$ remains monotonic up to the largest value of $U$ here
considered.

Results for the transversal spin current $J^s_y$ are shown in
Fig.~\ref{fig8}(d) for $V_{SO}/t=0.5$ and various values of $U$
on the $50\times 8$ scattering region. First, notice that the 
nonmonotonic behavior observed for $U=0$ is still present up to $U=4$,
surviving the large error bars of the calculation. Again, the most
interesting behavior is the enhancement of $J^s_y$ with $U$,
particularly in the region near the maximum point, For $U=4$, there is
a suppression of $J^s_y$ at small values of $\beta/\alpha$. This
suppression could be traced to the behavior of $J^s_{SO,y}$ acquiring
a {\em opposite} sign to that of $J^s_{hop,y}$ for this value of $U$.
This overall enhancement of $J^s_y$ with $U$ for $W=8$ complements the
result for $W=2$ obtained with td-DMRG shown in Fig.~\ref{fig5}(d),

\begin{figure}[t]
\includegraphics[width=0.9\columnwidth,angle=0]{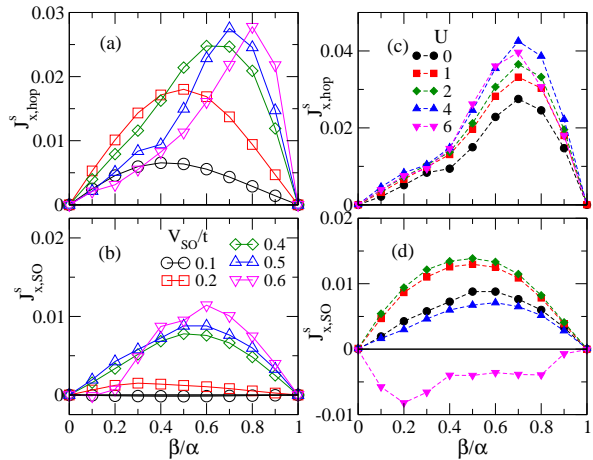}
\caption{(Color online) Longitudinal spin current on the $50\times 8$
cluster, as a function of $\beta/\alpha$. (a) Hopping contribution
and (b) SO contribution for various values of $V_{SO}/t$ as indicated 
on the plot. (c) Hopping contribution and (d) SO contribution for 
$V_{SO}/t=0.5$ and various values of $U$ indicated on the plot.
Results obtained by Landauer-Buttiker calculation.}
\label{fig9}
\end{figure}

To end this section, let us study the longitudinal polarized spin
current $J^{s}_x$ on the $50\times 8$ central region. The hopping,
$J^{s}_{hop,x}$, and SO, $J^{s}_{SO,x}$, contributions are shown
in Figs.~\ref{fig9}(a) and (b) respectively, as a function of
$\beta/\alpha$ and for various values of $V_{SO}/t$, $U=0$. As
argued before, $J^{s}_x$, as well as its two contributions,
$J^{s}_{hop,x}$ and $J^{s}_{SO,x}$, vanish at the pure Rashba
point and at the PSH point. Between these two limits, both
contributions to $J^{s}_x$ acquire finite values. It can be seen in
Fig.~\ref{fig9}(a) that $J^{s}_{hop,x}$, for a given value of
$V_{SO}/t$, follows qualitatively the same behavior as
$\sigma^s_{xx}$ for the same strip width $W=8$ (Fig.~\ref{fig3}(c)),
that is, it varies smoothly for small values of $V_{SO}/t$, and
then it develops a more pronounced maximum for larger $V_{SO}/t$.
This maximum becomes a peak for $V_{SO}/t\ge 0.5$, separating two
regions with clearly different behaviors. In addition, the position
of this maximum, $R^*$, shifts to higher $\beta/\alpha$ as $V_{SO}/t$
increases. Moreover, it should be noticed that $R^*$ coincides
with the corresponding one of the transversal spin current for the
same value of $V_{SO}/t$, as shown in Fig.~\ref{fig8}, and again this
is consistent with the behavior of $\sigma^s_{xx}$ and $\sigma_{sH}$,
where the values of $R^*$ were also coincident.

In contrast, as shown in Fig.~\ref{fig9}(b), the SO part of $J^{s}_x$
is much smaller than the hopping part, and its behavior is 
smoother. Hence, the more relevant contribution to $J^{s}_x$ is
of the hopping type, and this result is consistent with the linear
response result indicating that the most relevant contribution to the 
longitudinal spin conductivity is due to the correlation between
$\hat{j}_{SO,x}$ and $\hat{j}^s_{hop,x}$.

It is also important to notice that, as shown in Fig,~\ref{fig9}(c),
similarly to what was found for the spin accumulation and the
transversal spin current (Figs.~\ref{fig8}(c) and \ref{fig8}(d)),
$J^{s}_{hop,x}$
is {\em enhanced} by the Hubbard repulsion $U$, up to $U=4$. Besides,
$U$ preserves the overall dependence of $J^{s}_{hop,x}$ with
$\beta/\alpha$. In contrast, $J^{s}_{SO,x}$, shown
in Fig,~\ref{fig9}(d) is enhanced up to $U=2$ and becomes suppressed
by larger values of $U$, even becoming negative for $U=6$, that is
acquiring an opposite direction to $J^{s}_{hop,x}$.

\section{Conclusions}

In this work, three different techniques that cover different 
equilibrium and out-of equilibrium regimes, interacting and 
noninteracting systems, and ranges of strip widths, were employed to
study magneto-transport properties due to combined Rashba and
Dresselhaus spin-orbit couplings.

The first main result is an unexpected nonmonotonic behavior as a
 function of
the ratio of Dresselhaus to Rashba couplings, $\beta/\alpha$, in the
spin Hall conductivity, calculated in linear response, and in the
related transversal spin current, $J^{s}_y$, calculated within the
td-DMRG and Landauer-Buttiker approaches. This nonmonotonic
behavior is characterized by the presence of a maximum that
separates regions with different curvatures. This maximum has the
characteristic of a peak for small strip widths $W$, evolving into
a cusp for larger $W$. A peak is also present in the longitudinal
spin conductivity, and it is located at the same value
of $\beta/\alpha$ as the peak in the spin Hall conductivity for
the corresponding values of the SOI strength and $W$. Again, a
maximum separating asymmetric regions with different behavior was
observed in the related longitudinal spin
current, $J^{s}_x$. Moreover, these maxima in
$J^{s}_x$ are located at the same value of $\beta/\alpha$ as the ones
of $J^{s}_y$ for the same set of parameters.
Notice that for this result to make sense, the strip width and the
ratio of SOI, $\beta/\alpha$, have to be independent variables, which
excludes the possibility of the strips to be defined by electrical
gates.


The second major result of the present effort is the enhancement of
the main physical properties related to the spin Hall effect, the
spin accumulation and the transversal spin current, as well as the
longitudinal spin current, under the application of a Hubbard 
on-site repulsion with coupling $U$. This enhancement is present up
to a relatively high value of $U$, depending on the physical property,
beyond which this property saturates or starts to decrease.

To have sizable electron correlations, the main candidate
materials for building the strips could be 
transition metal oxides, such as SrTiO$_3$, mentioned in the 
Introduction, where the presence of SOI is ubiquitous. In addition,
strips of arbitrarily width could be edged from the surfaces or
interfaces involving these compounds with orthorhombic structure.
Of course, the region at the surfaces or interfaces where itinerant
SO processes take place is not a mathematical plane but it has a
finite depth, and it is not trivial to determine to what extent the
electron correlations due to d-orbitals are significant nor if 
the density of carriers is large enough for such correlations to have
some effect (on magnetic properties for instance). Although a full
investigation of the influence of the electron filling is out of the
scope of the present work, it would be important to study other
compounds where electron correlations, large spin-orbit coupling and
finite carrier density could be present. One of these new compounds
could be the orthorhombic perovskite iridate, SrIrO$_3$, which is a 
three-dimensional semimetal, and where the spin Hall effect has been
observed \cite{patri2017}, although Rashba or Dresselhaus types of SOC
have not yet been identified. Another candidate material is
Sr$_3$Ir$_2$O$_7$ lying close to the metal-Mott insulator transition
and exhibiting weak metallicity \cite{liuPRB2014}.

Finally, together with the study of varying electron filling, future
effort should be devoted to correlate the presently shown behavior of
the spin Hall conductivity and spin currents, with the behavior of 
magnetic properties, as it was for instance performed for the 
Rashba-Hubbard model in the two-chain strip \cite{riera13}. For the
isotropic 2D system, a study relating the spin conductivities with
dynamical magnetic susceptibilityes was done in 
Ref.~\cite{erlingsson} but its extension to the presently
studied system is certainly out of the scope of the present work.

\begin{acknowledgments}
The author wishes to thank A. Greco, I. Hamad, and L. Lara, for
useful discussions, 
and C. Gazza for helping to understand and write the KWANT code.
The author is partially supported by the Consejo Nacional de
Investigaciones Cient\'ificas y T\'ecnicas (CONICET) of Argentina.
through CONICET-PIP No. 11220120100389CO.
\end{acknowledgments}

\appendix
\section{Rashba and Dresselhaus Hamiltonians}
\label{app:RDham}

The Rashba SO Hamiltonian in the square lattice is obtained from:
\begin{align}
H_R = \alpha \sum_{l} [&(c_{l+y,\uparrow}^\dagger
 c_{l+y,\downarrow}^\dagger) (i\sigma^x) \left( \begin{array}{c}
  c_{l,\uparrow}  \\
  c_{l,\downarrow}
  \end{array} \right) - \nonumber \\
&(c_{l+x,\uparrow}^\dagger c_{l+x,\downarrow}^\dagger (i\sigma^y)) 
\left( \begin{array}{c}
  c_{l,\uparrow}  \\
  c_{l,\downarrow}
  \end{array} \right) + H. c.]
\label{Rsq0}
\end{align}
The Dresselhaus SO Hamiltonian $H_D$ results from a similar expression
just by interchanging $c_{l+y,\sigma}^\dagger c_{l+y,\sigma}$ with
$c_{l+x,\sigma}^\dagger c_{l+x,\sigma}$.

Assuming translational invariance along the $x$-axis, the SO part of 
the Hamiltonian in momentum space can be written as:
\begin{eqnarray}
H_R+H_D=\sum_{\mathbf{k}}
\begin{pmatrix}
A & B  \\
B^* & A 
\end{pmatrix} \qquad
\end{eqnarray}
where $A(k_x)$ and $B(k_x)$ are $W\times W$ matrices.

\section{Current operators}
\label{app:curop}

From the hopping term of the Hamiltonian, $H_h$, the following charge
current operators are obtained \cite{hamad}:
\begin{eqnarray}
\hat{j}_{\sigma,l,\mu}= -it (c_{l+\mu,\sigma}^\dagger c_{l,\sigma}
                           - c_{l,\sigma}^\dagger c_{l+\mu,\sigma}),
\end{eqnarray}
$\mu = x,y$, and the following spin current operators:
\begin{eqnarray}
\hat{j}^{s}_{l,\mu}= \frac{t}{2}(\hat{j}_{\uparrow,l,\mu} -
                  \hat{j}_{\downarrow,l,\mu}).
\end{eqnarray}
From the SO terms of the Hamiltonian, $H_R$ and $H_D$, the following
charge current operators are derived:
\begin{align}
\hat{j}_{SO,l,x}&=\hat{j}_{SO,l,x}^{'} + \hat{j}_{SO,l,x}^{''}
\nonumber\\
\hat{j}_{SO,l,y}&=\hat{j}_{SO,l,y}^{'} + \hat{j}_{SO,l,y}^{''}
\nonumber
\end{align}
and the following spin current operators:
\begin{align}
\hat{j}_{SO,l,x}^{s}&=\frac{1}{2}(\hat{j}_{SO,l,x}^{'} -
                      \hat{j}_{SO,l,x}^{''})
\nonumber\\
\hat{j}_{SO,l,y}^{s}&=\frac{1}{2}(\hat{j}_{SO,l,y}^{'} -
                      \hat{j}_{SO,l,y}^{''})
\nonumber
\end{align}

where the spin-selected SO charge currents are:

\begin{align}
\hat{j}_{SO,l,x}^{'} \equiv
\hat{j}_{SO,\uparrow \rightarrow \downarrow,l,x}&=i (\hat{h}_{SO,2,l,x}
               -\hat{h}_{SO,2,l,-x})
\nonumber\\
\hat{j}_{SO,l,x}^{''} \equiv
\hat{j}_{SO,\downarrow \rightarrow \uparrow,l,x}&=i (\hat{h}_{SO,1,l,x}
               -\hat{h}_{SO,1,l,-x})
\nonumber\\
\hat{j}_{SO,l,y}^{'} \equiv
\hat{j}_{SO,\uparrow \rightarrow \downarrow,l,y}&=i (\hat{h}_{SO,4,l,y}
                -\hat{h}_{SO,4,l,-y})
\nonumber\\
\hat{j}_{SO,l,y}^{''} \equiv
\hat{j}_{SO,\downarrow \rightarrow \uparrow,l,y}&=i (\hat{h}_{SO,3,l,y}
                -\hat{h}_{SO,3,l,-y})
\nonumber
\end{align}

and the SO Hamiltonian terms are defined as:

\begin{align}
\hat{h}_{SO,1,l,-x}&= -(\alpha,\beta) c_{l,\downarrow}^\dagger
               c_{l+x,\uparrow}
\nonumber\\
\hat{h}_{SO,2,l,x}&= ~~(\alpha,\beta) c_{l+x,\downarrow}^\dagger
               c_{l,\uparrow}
\nonumber\\
\hat{h}_{SO,3,l,-y}&= -(\beta,\alpha) c_{l,\downarrow}^\dagger
               c_{l+y,\uparrow}
\nonumber\\
\hat{h}_{SO,4,l,y}&= ~~(\beta,\alpha) c_{l+y,\downarrow}^\dagger
               c_{l,\uparrow}
\nonumber\\
\hat{h}_{SO,2,l,-x}&= (\alpha,-\beta) c_{l,\uparrow}^\dagger
               c_{l+x,\downarrow}
\nonumber\\
\hat{h}_{SO,1,l,x}&= (-\alpha,\beta) c_{l+x,\uparrow}^\dagger
               c_{l,\downarrow}
\nonumber\\
\hat{h}_{SO,4,l,-y}&= (\beta,-\alpha) c_{l,\uparrow}^\dagger
               c_{l+y,\downarrow}
\nonumber\\
\hat{h}_{SO,3,l,y}&= (-\beta,\alpha) c_{l+y,\uparrow}^\dagger
               c_{l,\downarrow}
\nonumber
\end{align}.


\begin{thebibliography}{}
\bibitem{note} Actually, the energy of the scattering region was 
      varied between $E_F-0.04$, and $E_F+0.04$.
\end{thebibliography}
\end{document}